\documentclass[proceedings]{stacs}
\stacsheading{2010}{561-572}{Nancy, France}
\firstpageno{561}

\usepackage[latin2]{inputenc}
\usepackage[english]{babel}
\usepackage{url}
\usepackage{amssymb}
\usepackage{amsmath}
\usepackage{amsthm}
\usepackage{tabularx}
\usepackage{times}
\usepackage{graphicx}
\usepackage{comment}
\newcommand{\X}{\mathcal{X}}
\newcommand{\G}{\mathcal{G}}
\newcommand{\torso}{\textup{torso}}
\newcommand{\hck}{$(H,C,K)$}
\newcommand{\hclk}{$(H,C,\text{$\le$}K)$}
\begin{document}

\newtheorem{claim}[thm]{Claim}
\newtheorem{subclaim}[thm]{Subclaim}
\newtheorem{defin}[thm]{Definition}


\title[Treewidth reduction]{Treewidth reduction for constrained separation and bipartization problems}

\author[lab1]{D.Marx}{D\'aniel Marx}
\address[lab1]{Tel Aviv University}
\email{dmarx@cs.bme.hu}

\author[lab2]{B.O'Sullivan}{Barry O'Sullivan}
\author[lab2]{I.Razgon}{Igor Razgon}
\address[lab2]{Cork Constraint Computation Centre, University College Cork}
\email{{b.osullivan,i.razgon}@cs.ucc.ie}




\begin{abstract}
  We present a method for reducing the treewidth of a graph while
  preserving all the minimal $s-t$ separators. This technique turns
  out to be very useful for establishing the fixed-parameter tractability of
  constrained separation and bipartization problems.
  To demonstrate the power of this technique, we
  prove the fixed-parameter tractability of a number of well-known separation
  and bipartization problems
  with various additional restrictions (e.g., the vertices being removed
  from the graph form an independent set).
  These results answer a number of open questions in the area of
  parameterized complexity.
\end{abstract}
\maketitle
\section{Introduction}
Finding cuts and separators is a classical topic of combinatorial
optimization and in recent years there has been an increase in
interest in the fixed-parameter tractability of such problems
\cite{MarxTCS,1132573,DBLP:conf/iwpec/Guillemot08a,DBLP:conf/csr/Xiao08,DBLP:journals/eor/GuoHKNU08,MR2330167,DBLP:conf/wads/ChenLL07,marxrazgon-esa2009}.
Recall that a problem is {\em fixed-parameter tractable} (or \textsc{FPT}) with
respect to a parameter $k$ if it can be solved in time $f(k)\cdot n^{O(1)}$ for
some function $f(k)$ depending only on $k$
\cite{MR2001b:68042,MR2238686,MR2223196}. In typical parameterized
separation problems, the parameter $k$ is the size of the separator we
are looking for, thus fixed-parameter tractability with respect to
this parameter means that the combinatorial explosion is restricted to
the size of the separator, but otherwise the running time depends
polynomially on the size of the graph.

The main technical contribution of the present paper is a theorem
stating that given a graph $G$, two terminal vertices $s$ and $t$, and
a parameter $k$, we can compute in a \textsc{fpt}-time a graph
$G^*$ having its treewidth bounded by a function
of $k$ while (roughly speaking) preserving all the minimal $s-t$ separators of
size at most $k$. Combining this theorem with the well-known Courcelle's Theorem,
we obtain a powerful tool for proving the fixed parameter tractability of constrained
separation and bipartization problems. We demonstrate the power of the
methodology with the following results.

\begin{itemize}
\item We prove that the \textsc{minimum stable $s-t$ cut} problem (Is
  there an independent set $S$ of size at most $k$ whose removal
  separates $s$ and $t$?) is fixed-parameter tractable. This problem
  received some attention in the community. Our techniques allow us to
  prove various generalizations of this result very easily. First,
  instead of requiring that $S$ is independent, we can require that it
  induces a graph that belongs to a hereditary class $\G$; the problem
  remains \textsc{fpt}. Second, in the \textsc{multicut} problem a
  list of pairs of terminals are given $(s_1,t_1)$, $\dots$,
  $(s_\ell,t_\ell)$ and the solution $S$ has to be a set of at most $k$ vertices that
  induces a graph from $\G$ and separates $s_i$ from $t_i$ for every
  $i$. We show that this problem is \textsc{fpt} parameterized by $k$
  and $\ell$, which is a very strong generalization of previous
  results \cite{MarxTCS,DBLP:conf/csr/Xiao08}. Third, the results
  generalize to the \textsc{multicut-uncut} problem, where two sets
  $T_1$, $T_2$ of pairs of terminals are given, and $S$ has to
  separate every pair of  $T_1$ and {\em should not} separate any
  pair of $T_2$.

\item We prove that the \textsc{exact stable bipartization} problem (Is there
an independent set of size \emph{exactly} $k$ whose removal makes the graph bipartite?) is
fixed-parameter tractable (\textsc{fpt}) answering an open question posed in 2001 by D{\'{\i}}az
et al. \cite{MR1907021}. We establish this result by proving that
the \textsc{stable bipartization} problem (Is there an independent set of size
\emph{at most} $k$ whose removal makes the graph bipartite?) is \textsc{fpt},
answering an open question posed by Fernau \cite{demaine_et_al:DSP:2007:1254}.

\item We show that the \textsc{edge-induced vertex
    cut} (Are there at most $k$ edges such that the removal of their
  endpoints separates two given terminals $s$ and $t$?) is
  \textsc{fpt}, answering an open problem posed in 2007 by Samer
  \cite{demaine_et_al:DSP:2007:1254}.  The motivation behind this
  problem is described in~\cite{DBLP:journals/corr/abs-cs-0607109}.

\end{itemize}

  We believe that the above results nicely demonstrate the
  message of the paper.  Slightly changing the definition of a
  well-understood cut problem usually makes
  the problem NP-hard and determining the parameterized complexity of
  such variants directly is by no means obvious. On the other hand,
  using our techniques, the fixed-parameter tractability of many such
  problems can be shown with very little effort. Let us mention
  (without proofs) three more variants that can be treated in a
  similar way: (1) separate $s$ and $t$ by the deletion of at most $k$
  edges and at most $k$ vertices, (2) in a 2-colored graph, separate
  $s$ and $t$ by the deletion of at most $k$ black and at most $k$
  white vertices, (3) in a $k$-colored graph, separate $s$ and $t$ by
  the deletion of one vertex from each color class.

As the examples above show, our method leads to the solution of several independent problems;
it seems that the same combinatorial difficulty lies at the heart of
these problems. Our technique manages to overcome this difficulty and
it is expected to be of use for further problems of similar
flavor. Note that while designing \textsc{fpt}-time
algorithms for bounded-treewidth graphs and in particular the use of Courcelle's
Theorem is a fairly standard technique, we use this technique
for problems where there is no bound on the treewidth
of the graph appearing in the input.



(Multiterminal) cut problems
\cite{MarxTCS,DBLP:journals/eor/GuoHKNU08,MR2330167,DBLP:conf/wads/ChenLL07}
play a mysterious,
and not yet fully understood, role in the fixed-parameter tractability of
certain problems. Proving that \textsc{bipartization}
\cite{ReedSmithVetta-OddCycle}, \textsc{directed feedback vertex set}
\cite{DBLP:journals/jacm/ChenLLOR08}, and \textsc{almost 2-sat}
\cite{ROicalp} are \textsc{fpt} answered longstanding open questions,
and in each case the algorithm relies on a non-obvious use of
separators.  Furthermore, \textsc{edge multicut} has been observed to
be equivalent to \textsc{fuzzy cluster editing}, a correlation
clustering problem
\cite{DBLP:conf/mfcs/BodlaenderFHMPR08,DBLP:journals/tcs/DemaineEFI06,DBLP:journals/ml/BansalBC04}.
Thus aiming for a better understanding of separators in a
parameterized setting seems to be a fruitful direction of research.
Our results extend our understanding of separators by showing that
various additional constraints can be accommodated. It is important to
point out that our algorithm is very different from previous
parameterized algorithms for separation problems
\cite{MarxTCS,DBLP:journals/eor/GuoHKNU08,MR2330167,DBLP:conf/wads/ChenLL07}.
Those algorithms in the literature exploit certain nice properties
of separators, and hence it seems impossible to generalize them for
the problems we consider here. On the other hand, our approach is very
robust and, as demonstrated by our examples, it is able to handle many
variants.



The paper assumes the knowledge of the definition of treewidth and its
algorithmic use, including Courcelle's Theorem (see the surveys
\cite{DBLP:conf/wg/Bodlaender06,GroheLGA}).

\section{Treewidth Reduction}\label{sec:treewidth-reduction}
The main combinatorial result of the paper is presented in this
section. 
We start with some preliminary definitions. Two slightly
different notions of separation will be used in the paper:
\begin{definition}
We say that a set $S$ of vertices {\em separates} sets of vertices $A$ and $B$ if
no component of $G\setminus S$ contains vertices from both $A\setminus
S$ and $B\setminus S$.
If $s$ and $t$ are two distinct vertices of $G$, then an $s-t$ {\em separator}
is a set $S$ of vertices disjoint from $\{s,t\}$ such that $s$ and $t$
are in different components of $G\setminus S$.
\end{definition}

In particular, if $S$ separates $A$ and $B$, then $A\cap B\subseteq
S$. Furthermore, given a set $W$ of vertices, we say that a set $S$ of
vertices is a {\em balanced separator} of $W$ if $|W\cap C|\le |W|/2$
for every connected component $C$ of $G\setminus S$.  A {\em $k$-separator}
is a separator $S$ with $|S|=k$.  The treewidth of a graph is closely
connected with the existence of balanced separators:
\begin{lemma}[\cite{ree97}, {\cite[Section 11.2]{MR2238686}}]\label{lem:balanced}
\
\begin{enumerate}
\item If $G(V,E)$ has treewidth greater than $3k$, then there is a set
$W\subseteq V$ of size $2k+1$ having no balanced $k$-separator.
\item If $G(V,E)$ has treewidth at most $k$, then every $W\subseteq V$
has a balanced $(k+1)$-separator.
\end{enumerate}
\end{lemma}
Note that the contrapositive of (1) in Lemma~\ref{lem:balanced} says
that if every set $W$ of vertices has a balanced $k$-separator,
then the treewidth is at most $3k$. This observation, and the following simple extension,
will be convenient tools for showing that a
certain graph has low treewidth.

\begin{lemma}\label{multibalanced}
Let $G$ be a graph, $C_1$,$\dots$, $C_r$ subsets of vertices, and let
$C:=\bigcup_{i=1}^r C_i$. Suppose that every $W_i\subseteq C_i$ has a
balanced separator $S_i\subseteq C_i$ of size at most $w$. Then every
$W\subseteq C$ has a balanced separator $S \subseteq C$ of
size $wr$.
\end{lemma}

If we are interested in separators of a graph $G$ contained in a subset $C$ of
vertices, then each component of $G\setminus C$ (or the neighborhood
of each component in $C$) can be replaced by
a clique, since there is no way to disconnect these components with
separators in $C$. The notion of torso and
Proposition~\ref{prop:torsosep} formalize this concept.

\begin{definition} \label{torsodef}
Let $G$ be a graph and $C\subseteq V(G)$. The graph $\torso(G,C)$ has
vertex set $C$ and vertices $a,b\in C$ are connected by an edge if
$\{a,b\} \in E(G)$ or there is a path $P$ in $G$ connecting $a$ and $b$ whose internal
vertices are not in $C$.
\end{definition}

\begin{proposition}\label{prop:torsosep}
  Let $C_1\subseteq C_2$ be two subsets of vertices in $G$ and let
  $a,b\in C_1$ be two vertices. A set $S\subseteq C_1$ separates $a$ and
  $b$ in $\torso(G,C_1)$ if and only if $S$ separates these vertices
  in $\torso(G,C_2)$.
In particular, by setting $C_2=V(G)$, we get that $S\subseteq C_1$
separates $a$ and $b$ in $\torso(G,C_1)$ if and only if it separates
them in $G$.
\end{proposition}

Analogously to Lemma~\ref{multibalanced}, we can show that if we have
a treewidth bound on $\torso(G,C_i)$ for every $i$, then these bounds add up for
the union of the $C_i$'s.
\begin{lemma}\label{lem:multitorsobound}
  Let $G$ be a graph and $C_1$,$\dots$, $C_r$ be subsets of $V(G)$
  such that for every $1\le i\le r$, the treewidth of $\torso(G,C_i)$
  is at most $w$. Then the treewidth of $\torso(G,C)$ for
  $C:=\bigcup_{i=1}^r C_i$ is at most $3r(w+1)$.
\end{lemma}


If the minimum size of an $s-t$ separator is $\ell$, then the {\em
  excess} of an $s-t$ separator $S$ is $|S|-\ell$ (which is always
nonnegative).  Note that if $s$ and $t$ are adjacent, then no $s-t$
separator exists, and in this case we say that the minimum size of an
$s-t$ separator is $\infty$.  The aim of this section is to show that,
for every $k$, we can construct a set $C'$ covering all the $s-t$ separators of
size at most $k$ such that $\torso(G,C')$ has treewidth bounded by a
function of $k$. Equivalently, we can require that $C'$ covers every
$s-t$ separator of excess at most $e:=k-\ell$, where $\ell$ is the
minimum size of an $s-t$ separator.

If $X$ is a set of vertices, we denote by $\delta(X)$ the set of those
vertices in $V(G)\setminus X$ that are adjacent to at least one vertex
of $X$. The following result is folklore; it can be proved by a simple
application of the uncrossing technique (see the proof below) and it can
be deduced also from the observations of \cite{MR592081} on the
strongly connected components of the residual graph after solving a
flow problem.
\begin{lemma}\label{lem:sepsequence}
  Let $s,t$ be two vertices in graph $G$ such that the minimum size of
  an $s-t$ separator is $\ell$.
  Then there is a collection $\X=\{X_1,\dots, X_q\}$ of sets where $\{s\}\subseteq X_i
  \subseteq V(G)\setminus (\{t\}\cup \delta(\{t\}))$ ($1\le i \le q$),
  such that
\begin{enumerate}
\item $X_1\subset X_2 \subset \dots \subset X_q$,
\item $|\delta(X_i)|=\ell$ for every $1 \le i \le q$, and
\item every $s-t$ separator of size $\ell$ is a subset of $\bigcup_{i=1}^{q}\delta(X_i)$.
\end{enumerate}
Furthermore, such a collection $\X$ can be found in polynomial time.
\end{lemma}

\proof
Let $\X=\{X_1,\dots, X_q\}$ be a collection of sets such that (2) and
(3) holds. Let us choose the collection such that $q$ is the minimum
possible, and among such collections, $\sum_{i=1}^{q}|X_i|^2$ is the
maximum possible. We show that for every $i,j$, either
$X_i\subset X_j$ or $X_j\subset X_i$ holds, thus the sets can be ordered
such that (1) holds.

Suppose that neither $X_i\subset X_j$ nor $X_j\subset X_i$ holds for
some $i$ and $j$. We show that after replacing $X_i$ and $X_j$ in $\X$
with the two sets $X_i\cap X_j$ and $X_i\cup X_j$, properties (2) and
(3) still hold, and the resulting collection $\X'$ contradicts the
optimal choice of $\X$. The function $\delta$ is well-known to be
submodular, i.e.,
\[
|\delta(X_i)|+|\delta(X_j)|\ge |\delta(X_i\cap X_j)|+|\delta(X_i\cup X_j)|.
\]
Both $\delta(X_i\cap X_j)$ and
$\delta(X_i\cup X_j)$ are $s-t$ separators
(because both $X_i \cap X_j$ and $X_i \cup X_j$ contain $s$)
and hence have size at least $k$.
The left hand side
is $2\ell$, hence there is equality and $|\delta(X_i\cap
X_j)|=|\delta(X_i\cup X_j)|=\ell$ follows. This means that property (2) holds after
the replacement.  Observe that $\delta(X_i\cap X_j) \cup
\delta(X_i\cup X_j) \subseteq \delta(X_i)\cup \delta(X_j)$: any edge
that leaves $X_i\cap X_j$ or $X_i\cup X_j$ leaves either $X_i$ or
$X_j$. We show that there is equality here, implying that property (3)
remains true after the replacement. It is easy to see that
$\delta(X_i\cap X_j) \cap \delta(X_i\cup X_j) \subseteq
\delta(X_i)\cap \delta(X_j)$, hence we have
\[
|\delta(X_i\cap X_j) \cup \delta(X_i\cup
X_j)| = 2\ell-|\delta(X_i\cap X_j) \cap \delta(X_i\cup
X_j)|\ge 2\ell-|\delta(X_i)\cap \delta(X_j)|=|\delta(X_i)\cup \delta(X_j)|,
\]
showing the required equality.

If $X_i\cap X_j$ or $X_i\cup X_j$ was already present in $\X$, then
the replacement decreases the size of the collection, contradicting
the choice of $\X$. Otherwise, we have that $|X_i|^2+|X_j|^2<|X_i\cap
X_j|^2+|X_i\cup X_j|^2$
(to verify this, simply represent $|X_i|$ as $|X_i \cap X_j|+|X_i \setminus X_j|$,
$|X_j|$ as $|X_i \cap X_j|+|X_j \setminus X_i|$, $|X_i\cup X_j|$ as
$|X_i \cap X_j|+|X_i \setminus X_j|+|X_j \setminus X_i|$ and do direct calculation
having in mind that both $|X_i \setminus X_j|$ and $|X_j \setminus X_i|$
are greater than $0$),
again contradicting the choice of $\X$.
Thus an optimal collection $\X$ satisfies (1) as well.

To construct $\X$ in polynomial time, we proceed as follows. It is
easy to check in polynomial time whether a vertex $v$ is in a minimum
$s-t$ separator, and if so to produce such a separator $S_v$. Let
$X_v$ be the set of vertices reachable from $s$ in $G\setminus
S_v$. It is clear that $X_v$ satisfies (2) and if we take the
collection $\X$ of all such $X_v$'s, then together they satisfy
(3). If (1) is not satisfied, then we start doing the replacements as
above. Each replacement either decreases the size of the collection or
increases $\sum_{i=1}^{t}|X_i|^2$ (without increasing the collection size),
thus the procedure terminates after a polynomial number of steps.
\qed

Lemma~\ref{lem:sepsequence} shows that the union
$C$ of all minimum $s-t$ separators can be covered by a chain of
minimum $s-t$ separators. It is not difficult to see that this chain
can be used to define a tree decomposition (in fact, a path
decomposition) of $\torso(G,C)$. This observation solves the problem
for $e=0$. For the general case, we use induction on $e$.

\begin{lemma}\label{lem:torsobound}
  Let $s,t$ be two vertices of graph $G$ and let $\ell$ be the minimum
  size of an $s-t$ separator. For some $e\ge 0$, let $C$ be the union
  of all minimal $s-t$ separators having \emph{excess} at most $e$
  (i.e. of size at most $k=\ell+e$).  Then, for some constant $d$,
   there is an $O(f(\ell,e)\cdot
  |V(G)|^d)$ time algorithm that returns a set $C'\supseteq C\cup \{s,t\}$ such
  that the treewidth of $\torso(G,C')$ is at most $g(\ell,e)$,
  where functions $f$ and $g$ depend only on $\ell$
  and $e$ .
\end{lemma}
\proof

We prove the lemma by induction on $e$.
Consider the collection $\X$ of Lemma~\ref{lem:sepsequence} and
define $S_i:=\delta(X_i)$ for $1\le i \le q$. For the sake of
uniformity, we define $X_0:=\emptyset$, $X_{q+1}:=V(G)\setminus
\{t\}$, $S_0:=\{s\}$, $S_{q+1}:=\{t\}$. For $1 \leq i \leq q+1$, let $L_i:=X_i\setminus
(X_{i-1} \cup S_{i-1})$.  Also, for $1 \le i \le q+1$ and two
disjoint \emph{non-empty} subsets $A,B$ of $S_i\cup S_{i-1}$, we define $G_{i,A,B}$ to
be the graph obtained from $G[L_i \cup A \cup B]$ by contracting the
set $A$ to a vertex $a$ and the set $B$ to a vertex $b$.
Taking into account that if $C$ includes a vertex of some $L_i$ then $e>0$,
we prove the key observation that makes it possible to use induction.

\begin{claim}\label{claim:layer}
If a vertex $v \in L_i$ is in $C$, then there are  disjoint non-empty
subsets $A,B$ of $S_i\cup S_{i-1}$ such that
$v$ is part of a minimal $a-b$ separator $K_2$ in $G_{i,A,B}$
of  size at most $k$ (recall that $k=\ell+e$) and excess at most $e-1$.
\end{claim}
\proof
By definition of $C$, there is a minimal $s-t$ separator $K$ of size at most $k$
  that contains $v$. Let $K_1:=K\setminus L_i$ and $K_2:=K\cap L_i$.
  Partition $(S_i\cup S_{i-1})\setminus K$ into the set $A$ of vertices
  reachable from $s$ in $G\setminus K$ and the set $B$ of vertices non-reachable
  from $s$ in $G\setminus K$.
Let us observe that both $A$ and $B$ are non-empty. Indeed,
  due to the minimality of $K$, $G$ has a path $P$ from $s$ to $t$
  such that $V(P) \cap K=\{v\}$. By selection of $v$, $S_{i-1}$ separates
  $v$ from $s$ and $S_i$ separates $v$ from $t$. Therefore, at least one
  vertex $u$ of $S_{i-1}$ occurs in $P$ before $v$ and at least one vertex $w$
  of $S_i$ occurs in $P$ after $v$. The prefix of $P$ ending at $u$ and the suffix
  of $P$ starting at $w$ are both subpaths in $G \setminus K$. It follows that
  $u$ is reachable from $s$ in $G \setminus K$, i.e. belongs to $A$ and that
  $w$ is reachable from $t$ in $G \setminus K$, hence non-reachable from $s$ and
  thus belongs to $B$.

  To see that $K_2$ is an $a-b$ separator in $G_{i,A,B}$, suppose that
  there is a path $P$ connecting $a$ and $b$ in $G_{i,A,B}$ avoiding
  $K_2$. Then there is a corresponding path $P'$ in $G$ connecting a
  vertex of $A$ and a vertex of $B$. Path $P'$ is disjoint from $K_1$
  (since it contains vertices of $L_i$ and $(S_i\cup
  S_{i-1})\setminus K$ only) and from $K_2$ (by
  construction). Thus a vertex of $B$ is reachable from $s$ in
  $G\setminus K$, a contradiction.

  To see that $K_2$ is a minimal $a-b$
  separator, suppose that there is a vertex $u\in K_2$ such that
  $K_2\setminus \{u\}$ is also an $a-b$ separator in $G_{i,A,B}$.
  Since $K$ is minimal, there is an $s-t$ path $P$ in $G\setminus
  (K\setminus u)$, which has to pass through $u$. Arguing as when
  we proved that $A$ and $B$ are non-empty, we observe that $P$ includes
  vertices of both $A$ and $B$, hence we can consider a minimal
  subpath $P'$ of $P$ between a vertex $a' \in A$ and a vertex $b' \in B$.
  We claim that all the internal vertices of $P'$ belong to $L_i$.
  Indeed, due to the minimality of $P'$, an internal vertex of $P'$
  can belong either to $L_i$ or to $V(G) \setminus (K_1 \cup L_i \cup S_{i-1} \cup S_i)$.
  If all the internal vertices of $P'$ are from the latter set then
  there is a path from $a'$ to $b'$ in $G \setminus (K_1 \cup L_i)$ and
  hence in $G \setminus (K_1 \cup K_2)$ in contradiction to $b' \in B$.
  If $P'$ contains internal vertices of both sets then $G$ has an edge
  $\{u,w\}$ where $u \in L_i$ while $w \in V(G) \setminus (K_1 \cup L_i \cup S_{i-1} \cup S_i)$.
  But this is impossible since $S_{i-1} \cup S_i$ separates $L_i$ from the rest
  of the graph. Thus it follows that indeed all the internal vertices
  of $P'$ belong to $L_i$. Consequently, $P'$ corresponds to a path in
  $G_{i,A,B}$ from $a$ to $b$ that avoids $K_2 \setminus u$, a contradiction
  that proves the minimality of $K_2$.

  Finally, we show that $K_2$ has excess at most $e-1$. Let
  $K'_2$ be a minimum $a-b$ separator in $G_{i,A,B}$.
  Observe that $K_1 \cup K'_2$ is an $s-t$ separator in $G$.
  Indeed, consider a path $P$ from $s$ to $t$ in $G \setminus (K_1 \cup K'_2)$.
  It necessarily contains a vertex $u \in K_2$, hence arguing as in the
  previous paragraph we notice that $P$ includes vertices of both $A$
  and $B$. Considering a minimal subpath $P'$ of $P$ between a vertex
  $a' \in A$ and $b' \in B$ we observe, analogously to the previous
  paragraph that all the internal vertices of this path belong to $L_i$.
  Hence this path corresponds to a path between $a$ and $b$ in $G_{i,A,B}$.
  It follows that $P'$, and hence $P$, includes a vertex of $K'_2$, a
  contradiction showing that $K_1 \cup K'_2$ is indeed an $s-t$ separator in $G$.
  Due to the minimality of $K_2$, $K'_2 \neq \emptyset$.
  Thus $K_1 \cup K'_2$ contains at least one vertex from
  $L_i$, implying that $K_1\cup K'_2$ is not a minimum $s-t$ separator
  in $G$. Thus $|K_2|-|K'_2|=(|K_1|+|K_2|)-(|K_1|+|K'_2|)<k-\ell=e$,
  as required.  This completes the proof of  Claim~\ref{claim:layer}. \qed

Now we define $C'$.
Let $C_0:=\bigcup_{i=0}^{q+1}S_i$. For $e=0$, $C'=C_0$.
Assume that $e>0$. For $1\le i \le q+1$ and
disjoint non-empty subsets $A,B$ of $S_i\cup S_{i-1}$. Let $C_{i,A,B}$ be
such a superset of the
union of all minimal $a-b$ separators of $G_{i,A,B}$ of size most $k$ and  excess
at most $e-1$ that $C_{i,A,B} \cup \{a,b\}$ satisfies
the induction assumption with respect to $G_{i,A,B}$
(if the minimum size of an $a-b$ separator of $G_{i,A,B}$ is greater than $k$
then we set $C_{i,A,B}=\emptyset$). We define $C'$ as the union of $C_0$ and all
sets $C_{i,A,B}$ as above. Observe that $C'$ is defined correctly in the sense
that any vertex $v$ participating in an $s-t$ minimal separator of size at most
$k$ indeed belongs to $C'$. For $e=0$, the correctness of $C'$ follows from the
definition of sets $S_i$. For $e>0$, the correctness  follows from the above
Claim if we take into account that since $\bigcup_{i=1}^{q+1} L_i \cup C_0=V(G)$,
$v$ belongs to some $L_i$.

We shall show that the treewidth of $\torso(G,C')$ is at most $g(\ell,e)$, a function
recursively defined as follows: $g(\ell,0):=6\ell$ and $g(\ell,e):=3\cdot (2\ell+3^{2\ell}\cdot
(g(\ell,e-1)+1))$ for $e>0$. We do this by showing that in graph $G$, every set $W \subseteq C'$
 has a balanced separator of size at most $2\ell$ (for $e=0$) and
at most $2\ell+3^{2\ell}\cdot (g(\ell,e-1)+1)$ (for $e>0$). By Proposition \ref{prop:torsosep},
this will imply that in $\torso(G,C')$, $W$ has a balanced separator with the same upper
bound. By Lemma~\ref{lem:balanced}(1), the desired upper bound on the treewidth will immediately
follow.

Let $W\subseteq C'$ be an arbitrary set. Let $1 \le i \le q+1$ be the
smallest value such that $|W\cap X_i|\ge |W|/2$.  Consider the
separator $S_i\cup S_{i-1}$ (whose size is at most $2\ell$).
In $G \setminus (S_i\cup S_{i-1})$, the sets $X_{i-1}$, $L_i$, and
$V(G) \setminus (S_i\cup S_{i-1} \cup X_{i-1} \cup L_i)$ are pairwise separated
from each other. By the selection of $i$, the first and the third sets do not
contain more than half of $W$. If $e=0$, then $C'$ is disjoint from $L_i$,
hence the treewidth upper bound follows for $e=0$. We assume that $e>0$ and, using the
induction assumption, will show that $W \cap L_i$ has a balanced separator $S$ of size
at most $3^{2\ell}\cdot (g(\ell,e-1)+1)$. This will immediately imply that $S \cup S_i \cup S_{i-1}$
is a balanced separator of $W$ of size at most $2\ell+3^{2\ell}\cdot (g(\ell,e-1)+1)$,
which, in turn, will imply the desired upper bound on the treewidth of
$\torso(G,C')$.

By the induction assumption, the treewidth of $\torso(G_{i,A,B}, C_{i,A,B})$ is
at most $g(\ell,e-1)$ for any pair of disjoint subsets $A$, $B$ of $S_i \cup S_{i-1}$
such that $G_{i,A,B}$ has an $a-b$ separator of size at most $k$. By
the combination of Lemma~\ref{lem:balanced}(2) and Proposition
\ref{prop:torsosep}, graph
$G$ has a balanced separator of size at most
$g(\ell,e-1)+1$ for any set $W_{i,A,B} \subseteq C_{i,A,B}$.
Let $C^*$ be the union of $C_{i,A,B}$ for all such $A$ and $B$.
Taking into account that the number of choices of $A$ and $B$
is at most $3^{2\ell}$, for any $W^* \subseteq C^*$, $G$ has a balanced
separator of size at most $3^{2\ell}\cdot (g(\ell,e-1)+1)$ according to
Lemma~\ref{multibalanced}. By definition of $C'$, $W \cap L_i \subseteq C^*$,
hence the existence of the desired separator $S$ follows.

We conclude the proof by showing that the above set $C'$ can be constructed in
time $O(f(\ell,e)\cdot |V(G)|^d)$.  In particular, we present
an algorithm whose running time is $O(f(\ell,e)\cdot (|V(G)|-2)^d)$ (we assume that $G$
has more than 2 vertices), where
$f(\ell,e)$ is recursively defined as follows: $f(\ell,0)=1$ and
$f(\ell,e)=f(\ell,e-1)\cdot 3^{2\ell}+1$ for $e>0$.

The set $X_i$ can be computed as shown in the proof of Lemma \ref{lem:sepsequence}.
Then the set $S_i$ can be obtained as in the first paragraph of the proof of the
present lemma. Their union results in $C_0$ which is $C'$ for $e=0$.
Thus for $e=0$, $C'$ can be computed in time $O(|V(G)|-2)^d)$ (instead of considering
$s$ and $t$, we may consider their sets of neighbors). Since the computation involves
computing a minimum cut, we may assume that $d>1$. Now assume that $e>0$.
For each $i$ such that $1 \leq i \leq q+1$ and $|L_i|>0$,
we explore all possible disjoint subsets $A$ and $B$ of $S_i \cup S_{i-1}$.
For the given choice, we check if the size of a minimum $a-b$ separator of $G_{i,A,B}$ is at most $k$
(observe that it can be done in $O(|L_i|^d)$) and if yes, compute the set $C_{i,A,B}$. By the induction
assumption, the computation takes $O(f(\ell,e-1) \cdot |L_i|^d)$. So, exploring all possible choices
of $A$ and $B$ takes $O(f(\ell,e-1) \cdot 3^{2\ell} \cdot |L_i|^d)$. The overall complexity of
computing $C'$ is \[O((|V(G)|-2)^d+f(\ell,e-1) \cdot 3^{2\ell} \cdot \sum_{i=1}^{q+1} |L_i|^d).\]
Since all $L_i$ are disjoint and $\bigcup_{i=1}^{q+1} L_i \subseteq V(G) \setminus \{s,t\}$,
$\sum_{i=1}^{q+1} |L_i| \leq |V(G)|-2$, hence $\sum_{i=1}^{q+1} (|L_i|)^d \leq (|V(G)|-2)^d$.
Taking into account the recursive expression for $f(\ell,e)$, the desired runtime follows.
\qed

\begin{remark}\label{rem:cube}
\textup{
The recursion $g(\ell,e):=3\cdot (2\ell+3^{2\ell}\cdot
g(\ell,e-1))$ implies that $g(\ell,e)$ is $2^{O(e\ell)}$, i.e., the
treewidth bound is exponential in $\ell$ and $e$. It is an obvious
question whether it is possible to improve this dependence to
polynomial. However, a simple example (graph $G$ is the $n$-dimensional
hypercube, $k=(n-1)n$, $s$ and $t$ are opposite vertices) shows that the function
$g(\ell,e)$ has to be exponential. The size of the
minimum $s-t$ separator is $\ell:=n$. We claim that every vertex $v$ of
the hypercube (other than $s$ and $t$) is part of a minimal $s-t$
separator of size at most $n(n-1)$. To see this, let $P$ be a shortest
path connecting $s$ and $v$. Let $P'=P-v$ be the subpath of $P$ connecting
$s$ with a neighbor $v'$ of $v$. Let $S$ be the neighborhood of $P'$;
clearly $S$ is an $s-t$ separator and  $v\in S$. However, $S\setminus
v$ is not an $s-t$ separator: the path $P$ is not blocked by
$S\setminus v$ as $S\setminus v$ does not contain any vertex farther
from $s$ than $v$.
Since $P'$ has at most $n-1$ vertices and every
vertex has degree $n$, we have $|S|\le n(n-1)$. Thus $v$ (and every
other vertex) is part of a
minimal separator of size at most $n(n-1)$. Hence if we set $\ell:=n$
and $e:=n(n-1)$, then $C$ contains every vertex of the hypercube. The
treewidth of an $n$-dimensional hypercube is $\Omega(2^n/\sqrt{n})$
\cite{MR2204113}, which is also a lower bound on $g(\ell,e)$.}
\end{remark}

The following theorem states our main combinatorial tool in a form that will be very convenient
to use.
\begin{theorem}[\textbf{The Treewidth Reduction Theorem}]
 \label{widthred} Let $G$ be a graph, $S \subseteq V(G)$, and let $k$ be
  an integer.  Let $C$ be the set of all vertices of $G$ participating
  in a minimal $s-t$ cut of size at most $k$ for some $s,t\in S$. Then
  there is an \textsc{fpt} algorithm, parameterized by $k$ and $|S|$,
  that computes a graph $G^*$ having the following properties:
\begin{enumerate}
\item $C \cup S \subseteq V(G^*)$
\item For every $s,t\in S$, a set $K \subseteq V(G^*)$ with $|K| \leq
  k$ is a minimal $s-t$ separator of $G^*$ if and only if $K \subseteq
  C\cup S$ and $K$ is a minimal $s-t$ separator of $G$.
\item The treewidth of $G^*$ is at most $h(k,|S|)$ for some function $h$.
\item For any $K \subseteq C$, $G^*[K]$ is isomorphic to $G[K]$.
\end{enumerate}
\end{theorem}

\proof
For every $s,t\in S$ that can be separated by the removal of at most $k$ vertices,
the algorithm of Lemma~\ref{lem:torsobound}
computes a set $C'_{s,t}$ containing all the minimal $s-t$
separators of size at most $k$. By Lemma~\ref{lem:multitorsobound},
if $C'$ is the union of these at most $\binom{|S|}{2}$ sets, then
$G'=\torso(G,C')$ has treewidth bounded by a function of $k$ and
$|S|$. Note that $G'$ satisfies all the requirements
of the theorem except the last one: two vertices of $C'$ non-adjacent in $G$
may become adjacent in $G'$ (see Definition \ref{torsodef}). To fix this problem
we subdivide each edge $\{u,v\}$ of $G'$ such that $\{u,v\} \notin E(G)$ into two edges
with a vertex between them, and, to avoid selecting this vertex into a cut, we split it
into $k+1$ copies. In other words, for each edge $\{u,v\} \in E(G') \setminus E(G)$
we introduce $k+1$ new vertices $w_1,\dots, w_{k+1}$ and replace $\{u,v\}$ by
the set of edges
$\{\{u,w_1\},\dots, \{u,w_{k+1}\},\{w_1, v\}, \dots, \{w_{k+1},v\}\}$. Let $G^*$
be the resulting graph. It is not hard to check that $G^*$ satisfies all the properties
of the present theorem.
\qed

\begin{remark}
\textup{The treewidth of $G^*$ may be larger than the treewidth of $G$.
We use the phrase ``treewidth reduction'' in the sense that the treewidth of $G^*$
is bounded by a function of $k$ and $|S|$, while the treewidth of $G$
is unbounded.
}
\end{remark}
\section{Constrained Separation Problems}
Let $\G$ be a class of graphs. Given a graph
$G$, vertices $s$ and $t$, and parameter $k$,
the $\G$-\textsc{mincut} problem asks if $G$
has an $s-t$ separator $C$ of size at most $k$ such that
$G[C] \in \G$. The following theorem is the central
result of this section.

\begin{theorem} \label{hereditary}
Assume that $\G$ is \emph{decidable} and \emph{hereditary} (i.e. whenever
$G \in \G$ then for any $V' \subseteq V$, $G[V'] \in \G$). Then
the $\G$-\textsc{mincut} problem is \textsc{fpt}.
\end{theorem}

\proof (Sketch)
Let $G^*$ be a graph satisfying the requirements of Theorem
\ref{widthred} for $S=\{s,t\}$.
According to Theorem \ref{widthred}, $G^*$ can be computed in \textsc{fpt} time.
We claim that $(G,s,t,k)$ is a `YES' instance of the $\G$-\textsc{mincut}
problem if and only if $(G^*,s,t,k)$ is a `YES' instance of this problem.
Indeed, let $K$ be an $s-t$ separator in $G$ such that $|K| \leq k$
and $G(K) \in \G$. Since $\G$ is hereditary, we may assume that $K$ is
minimal (otherwise we may consider a minimal subset of $K$ separating $s$ from
$t$). By the second and fourth properties of $G^*$ (see Theorem \ref{widthred}),
$K$ separates $s$ from $t$ in $G^*$ and $G^*[K] \in \G$. The opposite
direction can be proved similarly.

Thus we have established an \textsc{fpt}-time reduction from an instance of
the $\G$-\textsc{mincut} problem to another instance of this problem where the treewidth
is bounded by a function of parameter $k$. Now, let $G_1=(V(G^*),E(G^*),ST)$ be a labeled
graph where $ST=\{s,t\}$. We present an algorithm for constructing
a monadic second-order (\textsc{mso}) formula
$\varphi$ whose atomic predicates (besides equality) are $E(x_1,x_2)$ (showing that
$x_1$ and $x_2$ are adjacent in $G^*$) and predicates of the form $X(v)$ (showing
that $v$ is contained in $X \subseteq V$), whose size is bounded by a function
of $k$, and $G_1 \models \varphi$ if and only if $(G^*,s,t,k)$ is a `YES' instance
of the $\G$-\textsc{mincut} problem. According to a restricted version of
the well-known Courcelle's Theorem (see the survey article of Grohe \cite{GroheLGA},
Remarks 3.19\footnote{Although the branchwidth
of $G_1$ appears in the parameter, it can be replaced by the treewidth of $G_1$
since the former is bounded by a function of $k$ if and only
if the latter is \cite{obstruct}.} and 3.20),
it will follow that the $\G$-\textsc{mincut} problem is \textsc{fpt}.
The part of $\varphi$ describing the separation of $s$ and $t$ is based on the ideas from \cite{MR2330167}.

We construct the formula $\varphi$ as
\[\varphi=\exists C ( \text{AtMost}_k(C) \wedge \text{Separates}(C)
\wedge \text{Induces}_{\G}(C)),
\]
where $\text{AtMost}_k(C)$ is true if and only if $|C| \leq k$, $\text{Separates}(C)$ is true
if and only if $C$ separates the vertices of $ST$ in $G^*$, and
$\text{Induces}_{\G}(C)$ is true if and only $C$ induces a graph of $\G$.

In particular, $\text{AtMost}_k(C)$ states that $C$ does not have $k+1$ mutually non-equal elements:
this can be implemented as\[
\forall c_1, \dots, \forall c_{k+1} \bigvee_{1 \leq i,j \leq k+1} (c_i=c_j).\]

Formula $\text{Separates}(C)$ is a slightly modified formula $\text{uvmc}(X)$ from \cite{MR2330167}, that looks as follows:
\begin{multline*}
\forall s \forall t \forall Z \big(ST(s) \wedge ST(t) \wedge \neg (s=t) \wedge
\neg C(s) \wedge \neg C(t) \wedge \text{Connects}(Z,s,t)\big )
\rightarrow \big(\exists v (C(v) \wedge Z(v)) )\big),
\end{multline*}

where  $\text{Connects}(Z,s,t)$ is true if and only if in the modeling graph
there is a path from $s$ and $t$ all vertices
of which belong to $Z$. For the definition of the predicate $\text{Connects}$, see Definition 3.1 in
\cite{MR2330167}.

To construct $\text{Induces}_{\G}(C)$, we explore all possible graphs having at most $k$ vertices
and for each of these graphs we check whether it belongs to $\G$. Since the number of graphs being
explored depends on $k$ and $\G$ is a decidable class, in \textsc{fpt} time we can compile
the set $\{G'_1, \dots, G'_r\}$ of all graphs of at most $k$ vertices that belong to $\G$.
Let $k_1, \dots k_r$ be the respective numbers of vertices of $G'_1,\dots G'_r$.
Then $\text{Induces}_{\G}(C)=\text{Induces}_1(C) \vee \dots \vee \text{Induces}_r(C)$, where $\text{Induces}_i(C)$
states that $C$ induces $G'_i$. To define $\text{Induces}_i$, let $v_1,\dots, v_{k_i}$ be the set
of vertices of $G'_i$ and define $\text{Adj}_i(c_1, \dots, c_{k_i})$ as the conjunction of
all $E(c_x,c_y)$ such that $v_x$ and $v_y$ are adjacent in $G'_i$ and of all
$\neg E(c_x,c_y)$ such that $v_x$ and $v_y$ are not adjacent in $G'_i$.
Then
\begin{multline*}
\text{Induces}_i(C)= \text{AtMost}_{k_i}(C) \wedge \exists c_1 \dots \exists c_{k_i}
\Big(\bigwedge_{1 \leq j \leq k_i} C(c_j) \wedge \bigwedge_{1 \leq x,y \leq k_i} c_x \neq c_y \wedge \text{Adj}_i(c_1, \dots, c_{k_i})\Big).
\end{multline*}

It is not hard to verify that indeed $G_1 \models \varphi$ if and only if $(G^*,s,t,k)$ is a `YES' instance
of the $\G$-\textsc{mincut} problem.
\qed

 In particular, let $\G^0$ be the class of all graphs without edges. Then
$\G^0$-\textsc{mincut} is the \textsc{minimum stable $s-t$ cut} problem whose fixed-parameter tractability
has been posed as an open question by Kanj \cite{KanjPersonal}. Clearly, $\G^0$ is hereditary
and hence the $\G^0$-\textsc{mincut} is \textsc{FPT}.

Theorem~\ref{hereditary} can be used to decide if there is an $s-t$
separator of size {\em at most} $k$ having a certain property, but
cannot be used if we are looking for $s-t$ separators of size {\em
  exactly $k$.} We show (with a very easy argument) that some of these
problems actually become hard if the size is required to be exactly
$k$. Let graph $G'$ be obtained from graph $G$ by introducing two
isolated vertices $s$ and $t$. Now there is an independent set of size
exactly $k$ that is an $s-t$ separator in $G'$ if and only if there is
an independent set of size $k$ in $G$, implying that finding such a
separator is W[1]-hard.

\begin{theorem}\label{th:exacthardness}
  It is \textup{W[1]}-hard 
to decide if $G$ has an $s-t$
  separator that is an independent set of size exactly $k$.
\end{theorem}

Samer and Szeider \cite{DBLP:journals/corr/abs-cs-0607109} introduced the notion of {\em edge-induced
  vertex-cut} and the corresponding computational problem: given a graph $G$ and two
vertices $s$ and $t$, the task is to decide if there are $k$ edges such that deleting
the {\em endpoints} of these edges separates $s$ and $t$. It remained
an open question in \cite{DBLP:journals/corr/abs-cs-0607109} whether this problem is
\textsc{fpt}. Samer reposted this problem as an open question in \cite{demaine_et_al:DSP:2007:1254}.
Using Theorem \ref{hereditary}, we answer this question positively.
For this purpose, we introduce $\G_k$, the class of graphs where the number of vertices
minus the size of the maximum matching is at most $k$, observe that this class is hereditary,
and show that $(G,s,t,k)$ is a `YES'-instance of the {\em edge-induced vertex-cut} problem
if and only if $(G,s,t,2k)$ is a `YES' instance of the $\G_k$-mincut problem. Then we apply Theorem \ref{hereditary}
to get the following corollary.

\begin{corollary} \label{edgeinduced}
The \textsc{edge-induced vertex-cut} problem
is \textsc{fpt}.
\end{corollary}


\textsc{multicut} is the generalization of \textsc{mincut} where,
instead of $s$ and $t$, the input contains a set $(s_1,t_1)$, $\dots$,
$(s_\ell,t_\ell)$ of terminal pairs. The task is to find a set $S$
of at most $k$ nonterminal vertices that separate $s_i$ and $t_i$ for
every $1\le i \le \ell$. \textsc{multicut} is known to be \textsc{fpt}
\cite{MarxTCS,DBLP:conf/csr/Xiao08} parameterized by $k$ and $\ell$.
In the \textsc{$\G$-multicut} problem, we additionally require that
$S$ induces a graph from $\G$. It is not difficult to generalize
Theorem~\ref{hereditary} for \textsc{$\G$-multicut}: all we need to do
is to change the construction of $\varphi$ such that it requires the
separation of each pair $(s_i,t_i)$.
We state this here in an even more general form.
In the $\G$-\textsc{multicut-uncut} problem  the input contains an
additional integer $\ell' \leq \ell$, and we change the problem by requiring for
every $\ell' \le i \le \ell$ that $S$ {\em does not} separate $s_i$
and $t_i$.

\begin{theorem} \label{hermultiuncut}
If $\G$ is \emph{decidable} and \emph{hereditary,} then $\G$-\textsc{multicut-uncut} is
\textsc{fpt} parameterized by $k$
and $\ell$.
\end{theorem}
Theorem~\ref{hermultiuncut} helps clarify a theoretical issue. In
Section~\ref{sec:treewidth-reduction}, we defined $C$ as the set of
all vertices appearing in minimal $s-t$ separators of size at most $k$.
There is no obvious way of finding this set in \textsc{fpt}-time and
Lemma~\ref{lem:multitorsobound} produces only a superset $C'$ of
$C$. However, Theorem~\ref{hermultiuncut} can be used to find $C$: a
vertex $v$ is in $C$ if and only if there is a set $S$ of size at most $k-1$ and
two neighbors $v_1,v_2$ of $v$ such that $S$ separates $s$ and $t$ in
$G\setminus v$, but $S$ does not separate $s$ from $v_1$ and  $t$ from
$v_2$ in $G\setminus v$ (including the possibility that $v_1=s$ or
 $v_2=t$).

\section{Constrained Bipartization Problems}\label{sec:bipartite-deletion}
Reed et al.~\cite{ReedSmithVetta-OddCycle} solved a longstanding open
question by proving the fixed-parameter tractability of the
\textsc{bipartization} problem: given a graph $G$ and an integer $k$,
find a set $S$ of at most $k$ vertices such that $G\setminus S$ is
bipartite (see also \cite{iwoca-oct} for a somewhat simpler
presentation of the algorithm). In fact, they showed that the
\textsc{bipartization} problem can be solved by at most $3^k$
applications of a procedure solving \textsc{mincut}. The key result
that allows to transform \textsc{bipartization} to a separation
problem is the following lemma.

\begin{lemma}\label{lem:bipartitesep}
Let $G$ be a bipartite graph and let $(B',W')$ be a 2-coloring of the
vertices.
Let $B$ and $W$ be two subsets of $V(G)$.  Then for any $S$, $G
\setminus S$ has a 2-coloring where $B \setminus S$ is black and $W
\setminus S$ is white if and only if $S$ separates $X:=(B\cap B')\cup
(W\cap W')$ and $Y:=(B\cap W')\cup (W\cap B')$.
\end{lemma}

In this section we consider the $\G$-\textsc{bipartization}
problem: a generalization of the \textsc{bipartization}
problem where, in addition to $G\setminus S$ being bipartite, it is
also required that $S$ induces a graph belonging to a class $\G$.

\begin{theorem}\label{th:genbipartite}
$\G$-\textsc{bipartization} is \textsc{fpt} if $\G$ is hereditary
and decidable.
\end{theorem}
\proof  Using the algorithm of \cite{ReedSmithVetta-OddCycle}, we first try to find a set $S_0$ of size at most
  $k$ such that $G\setminus S_0$ is bipartite. If no such set exists,
  then clearly there is no set $S$ satisfying the requirements.
  Otherwise, we branch in $3^{|S_0|}$ directions: each vertex of $S_0$
  is removed or colored black or colored white.
  For a particular branch, let $R=\{v_1, \dots, v_r\}$ be the vertices of
  $S_0$ to be removed and let $B_0$ (resp., $W_0$) be the vertices
  of $S_0$ having color black (resp., white) in a 2-coloring of
  the resulting bipartite graph. Let us call a set $S$ such that
  $S \cap S_0=R$, and $G \setminus S$ is bipartite and having a 2-coloring where $B_0$ and $W_0$ are colored black and white, respectively,
  a set \emph{compatible} with $(R,B_0,W_0)$. Clearly, $(G,k)$ is a `YES'
  instance of the $\G$-\textsc{bipartization} problem if and only if for at
  least one branch corresponding to partition $(R,B_0,W_0)$ of $S_0$, there is
  a set compatible with $(R,B_0,W_0)$ having size at most $k$ and such
  that $G[S] \in \G$. Clearly, we need to check only those
  branches where $G[B_0]$ and $G[W_0]$ are both independent sets.

  We transform the problem of finding a set compatible with $(R,B_0,W_0)$ into a
  separation problem. Let $(B',W')$ be a 2-coloring of $G\setminus
  S_0$. Let $B=N(W_0)\setminus S_0$ and $W=N(B_0)\setminus S_0$.  Let
  us define $X$ and $Y$ as in Lemma~\ref{lem:bipartitesep}, i.e.,
  $X:=(B\cap B')\cup (W\cap W')$, and $Y:=(B\cap W')\cup (W\cap B')$.
  We construct a graph $G'$ that is obtained from $G$ by deleting the
  set $B_0\cup W_0$, adding a new vertex $s$ adjacent to $X\cup R$,
  and adding a new vertex $t$ adjacent with $Y\cup R$. Note that every
  $s-t$ separator in $G'$ contains $R$. By
  Lemma~\ref{lem:bipartitesep}, a set $S$ is compatible with
  $(R,B_0,W_0)$ if and only if $S$ is an $s-t$ separator in $G'$. Thus
  what we have to decide is whether there is an $s-t$ separator $S$ of
  size at most $k$ such that $G'[S]=G[S]$ is in $\G$. That is, we have
  to solve the $\G$-\textsc{mincut} instance $(G',s,t,k)$. The
  fixed-parameter tractability of the $\G$-\textsc{bipartization}
  problem now immediately follows from Theorem \ref{hereditary}.
\qed

Theorem \ref{th:genbipartite} immediately implies that the
\textsc{stable bipartization} problem is \textsc{fpt}: just set $\G$
to be the class of all graphs without edges. This answers an open
question of Fernau~\cite{demaine_et_al:DSP:2007:1254}.  Next, we show
that the \textsc{exact stable bipartization} problem is \textsc{fpt},
answering a question posed by D{\'{\i}}az et al. \cite{MR1907021}.
This result may seem surprising because the corresponding exact separation
problem is W[1]-hard by Theorem \ref{th:exacthardness} and hence the
approach of Theorem~\ref{th:genbipartite} is unlikely to work.
Instead, we argue that under appropriate conditions, any solution of
size at most $k$ can be extended to an independent set of size exactly
$k$.

\begin{theorem}\label{th:bipartiteindepexact}
Given a graph $G$ and an integer $k$, deciding whether $G$ can be made
bipartite by the deletion of an independent set of size exactly $k$ is
fixed-parameter tractable.
\end{theorem}
\proof  (Sketch)
It is more convenient to consider an annotated version of the problem
where the independent set being deleted has to be a subset of a set $D \subseteq V(G)$
given as part of the input. Without the annotation, $D$ is initially set
to $V(G)$. If $G$ is not bipartite, then the algorithm starts by
finding an odd cycle $C$ of minimum length (which can be done in
polynomial time). It is not difficult to see that the minimality of
$C$ implies that either $C$ is a
triangle or $C$ is chordless. Moreover, in the latter case,
every vertex not in $C$ is adjacent to at most 2 vertices
of the cycle.

If $|V(C) \cap D|=0$, then clearly no subset of $D$ is a solution.
If $1\le |V(C) \cap D| \leq 3k+1$, then we branch on the selection of each vertex $v \in V(C) \cap D$
into the set $S$ of vertices being removed and apply the algorithm recursively
with the parameter $k$ being decreased by $1$ and the set $D$ being updated by the removal
of $v$ and $N(v) \cap D$. If $|V(C) \cap D| > 3k+1$, then we apply the approach of
Theorem \ref{th:genbipartite} to find an independent set $S\subseteq D$ of size at most $k$
whose removal makes the graph bipartite, and then argue that $S$ can
be extended to an independent set of size exactly $k$. To ensure that $S \subseteq D$, we may,
for example split all vertices $v \in V(G) \setminus D$ into $k+1$ independent copies with
the same neighborhood as $v$. If $|S|=k$, we are done. Otherwise, $|S|=k'<k$. In this case
we observe that by the minimality of $C$,  each vertex of $S$ (either in $C$ or
outside $C$) forbids the selection of at most $3$
vertices of $V(C) \cap D$ including itself. Thus the number of vertices of $V(C) \cap D$ allowed
for selection is at least $3k+1-3k'=3(k-k')+1$. Since the cycle is chordless, we can select
$k-k'$ independent vertices among them and thus complement $S$ to be of size exactly $k$.

The above algorithm has a number of stopping conditions, the only non-trivial of them occurs
if $G$ is bipartite but $k>0$. In this case we check if $G[D]$ has $k$ independent
vertices, which can be done in a polynomial time.
\qed

\section*{Acknowledgements}
The research of D\'aniel Marx was supported by ERC Advanced Grant DMMCA.
The research of Barry O'Sullivan and Igor Razgon was supported by Science Foundation Ireland through Grant 05/IN/I886.
We would like to thank the anonymous referees for spotting a number of minor mistakes in the preliminary version
of this paper. Fixing those mistakes in the camera-ready version allowed us to significantly improve its quality.


{\footnotesize 
\bibliographystyle{abbrv}
\bibliography{Marx} 
}
\vspace{-1.2cm}

\end{document}